# A large area, high counting rate micromegas-based neutron detector for BNCT

Zhujun Fang[1,2], Zhiyong Zhang[1,2,*], Bin Shi[3], Wei Jiang[1,2], Xianke Liu[1,2], Siqi He[1,2], Jun Chen[3], Ping Cao[1,2], Jianbei Liu[1,2], Yi Zhou[1,2], Ming Shao[1,2], Botian Qu[3], Shufeng Zhang[3], Qian Wang[3]

1. State Key Laboratory of Particle Detection and Electronics, University of Science and Technology of China, Hefei 230026, China
2. Department of Modern Physics, University of Science and Technology of China, Hefei 230026, China
3. National Key Laboratory for Metrology and Calibration Techniques, China Institute of Atomic Energy, Beijing, 102413, China
* Corresponding author, e-mail: zhzhy@ustc.edu.cn

Abstract: Beam monitoring and evaluation are very important to boron neutron capture therapy (BNCT), and a variety of detectors have been developed for these applications. However, most of the detectors used in BNCT only have a small detection area, leading to the inconvenience of the full-scale 2-D measurement of the beam. Based on micromegas technology, we designed a neutron detector with large detection area and high counting rate. This detector has a detection area of 288 mm × 288 mm and can measure thermal, epithermal, and fast neutrons with different detector settings. The BNCT experiments demonstrated that this detector has a very good 2-D imaging performance for the thermal, epithermal, fast neutron and gamma components, a highest counting rate of 94 kHz/channel, and a good linearity response to the beam power. Additionally, the flux fraction of each component can be calculated based on the measurement results. The Am-Be neutron source experiment indicates that this detector has a spatial resolution of approximately 1.4 mm, meeting the requirements of applications in BNCT. It is evident that this micromegas-based neutron detector with a large area and high counting rate capability has great development prospects in BNCT beam monitoring and evaluation applications.
Key words: Micromegas-based neutron detector, BNCT, large area, high counting rate, 2-D imaging.

1. Introduction:

Boron neutron capture therapy (BNCT) is a new treatment for some cancers [1]. The designed peak energy of the neutron beam used in BNCT is usually in the thermal neutron region or epithermal neutron region. However, the real neutron energy ranges from the thermal region to the fast region due to the limitation of the neutron moderator. It is very important to measure and monitor the flux, range and energy of neutron beams in BNCT treatment, which is of great significance to improve the curative effect and reduce the irradiation dose of patients [2]. Several kinds of neutron detectors have been developed for BNCT neutron beam monitoring. Scintillator detectors, such as Li-glass detectors, and semiconductor detectors, such as CdZnTe detectors, can be used to monitor the neutron beam dose and flux [3-5], as well as diamond detectors [6]. Additionally, activation detectors with various metal-activated tablets could be used to make an indirect measurement of the neutron flux [7]. In these detectors, most of the effective detection sizes are several mm to tens of mm, so a linear scan is required to make a full measurement of the neutron beam flux. Meanwhile, the spatial resolution is also in mm magnitude by the limit of the detector cell size. In this case, the uncertainty of the neutron beam will affect the measurement result, and it is quite complex to make a 2-D measurement of the neutron beam. Some researchers proposed a detector consisting of a Li-doped zinc plane as a neutron converter and a CCD camera as a scintillation detector [8]. This detector can measure the dose distribution in a water phantom, but with the disadvantage of high complexity. In these detector, the highest total counting rate is several MHz to meet the measurement demands in 100% neutron source power [3]. It is another requirement for the neutron detector used in BNCT.

The micropattern gaseous detector (MPGD) could effectively realize a large detection area, sub mm spatial resolution, and high counting rate capability, such as the GEM, THGEM, MM detectors and so on [9-11]. A thermal bonding method for the micromegas manufacturing process has been developed by the researchers from University of Science and Technology of China, promoting thickness uniformity in the electron multiplication region and realizing large area mesh bonding with high robustness [12]. Based on this method, we performed a large-area neutron beam monitoring detector for BNCT. Additionally, a high counting rate capability can be achieved in this detector with specially designed electronic systems. The detector and electronic system are tested in the in-hospital neutron irradiator BNCT beam at Beijing Capture Technology Corporation [13]. This BNCT system has a thermal neutron beam and an epithermal neutron beam, having a beam outlet diameter of 12 cm, a highest thermal neutron flux of $1.9\times10^9$/cm$^2$/s and an epithermal neutron flux of $5.0\times10^8$/cm$^2$/s.

In this paper, the design and experimental evaluations of this large area high counting rate micromegas-based neutron detector will be introduced in detail, demonstrating good performance in BNCT beam monitoring and evaluation applications.

2. The design of the micromegas-based neutron detector

Figure 1 shows a schematic of the micromegas-based neutron detector with two kinds of settings. The first is the thermal neutron detector, consisting of the drift cathode, the $^{nat}$LiF thin film as the neutron converter, the 5 mm gas gap, the standard mesh, and the readout printed circuit board (PCB) plate with a 300-400 nm thick Ge coating [14]. The reason for using the $^{nat}$LiF film instead of the $^6$LiF film is that neutron beam flux can be as high as $10^9$ neutron/cm$^2$/s in the common BNCT facilities. Therefore, the use of $^6$LiF will result in a very high counting rate that exceeds the electronic measuring capability. The second is the fast neutron detector, removing the $^{nat}$LiF film and using a 3 mm thick working gas as the fast neutron recoil material. The width of the gas gap influences the fast neutron detection efficiency and should be adjusted according to the predicted BNCT flux. Both detectors have a detection area of 288 mm × 288 mm, and the pitch of the X and Y readout strips is 1.5 mm. The spacer height of the micromegas is 100 μm, as the conventional setting of the thermal bonding method [12]. A resistive Ge layer with a surface resistance of tens to hundreds of MΩ/□ is used to suppress sparking, and a point array fast grounding design is arranged to obtain a high counting rate capability [15]. As shown in Fig. 2. The grounded points have a diameter of 0.5 mm and a pitch of 10 mm.

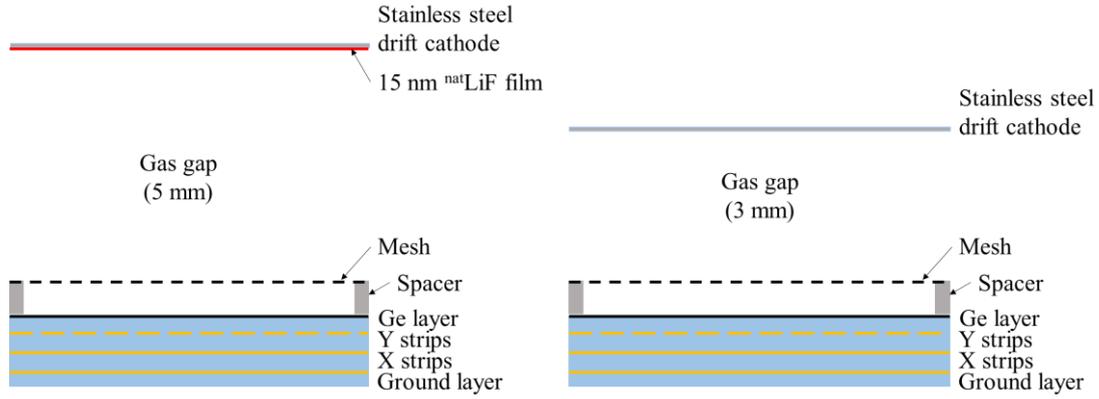

Fig. 1 (Left) Schematic of the thermal neutron detector. (Right) Schematic of the fast neutron detector.

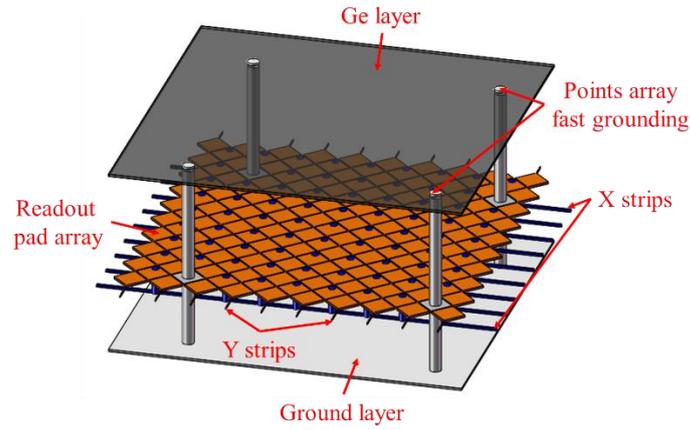

Fig. 2 The fast grounding design in the micromegas-based neutron detector. (Note: this figure is the schematic of the fast grounding point array connection and the readout pad array setting, the PCB substrate between these layers are undrawn.)

In the beam flux monitoring and evaluation applications of BNCT, the neutron detection efficiency should be determined according to the predicted flux and counting rate capability of the detector. Equation (1) presents the calculation of neutron detection efficiency, where σ is the neutron reaction cross-section, $N_A$ is the Avogadro's constant, ρ is the density, T is the thickness, and $M_0$ is the molecular weight. In the thermal neutron detector, the $^{nat}$LiF film has a thickness of 15 nm, a density of 2.64g/cm$^3$, a neutron reaction cross-section of 71.3 barn [16], indicating a detection efficiency of approximately 6.6×10$^{-6}$ for the 25.3 meV thermal neutron. In the fast neutron detector, the working gas component should be well considered to obtain a suitable detection efficiency and recoil energy. In this study, we use 3 mm thick Ar:$CO_2$ =93:7 as the working gas, and neutron scattering cross-section of carbon, oxygen, and argon is approximately 4.5 barn, 3.7 barn and 0.6 barn for fast neutrons, respectively [16]. Therefore, the detector has a detection efficiency of approximately 1.0×10$^{-5}$ for fast neutrons.

$$\varepsilon = 1 - e^{-\sigma \cdot N_A \cdot \rho \cdot T / M_0} \qquad \text{Equation (1)}$$

As shown in Fig. 3, the detector shell consists of an aluminum frame and a flat stainless steel cathode plane. This compact design can realize a very thin detector edge and small dead zone, as well as high reliability and robustness. In this design, the main differences between the thermal neutron detector and fast neutron detector are the presence or absence of the $^{nat}$LiF film and the thickness of the gas gap, so it is convenient to switch the detector mode by changing the drift cathode. For the convenience of the BNCT experiment, we manufactured a thermal neutron detector and a fast neutron detector.

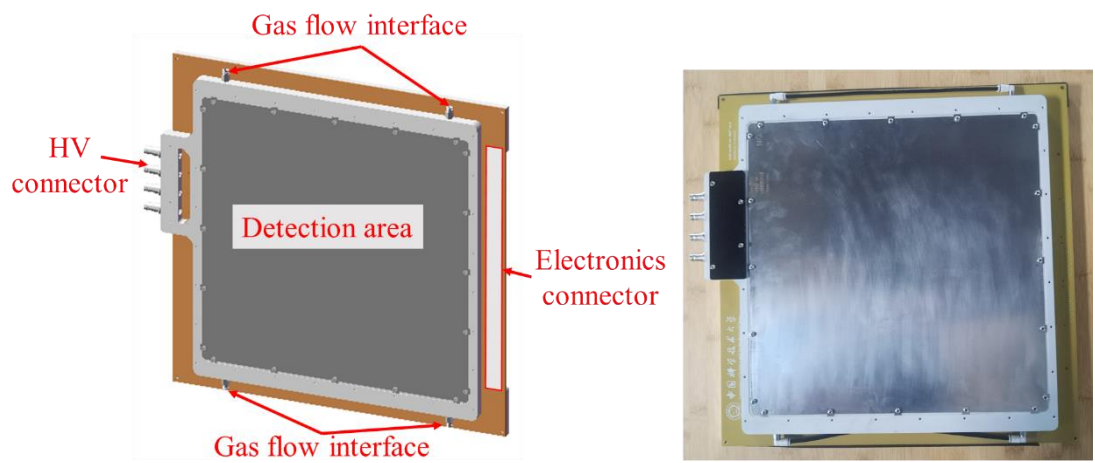

Fig. 3 (Left) The detector structure design. (Right) The manufactured micromegas-based neutron detector.

A dedicated readout electronics system featuring high counting rate for this detector is developed. The electronics system contains 12 data acquisition boards based on a PXIe platform, and each board holds 32 readout channels. A single readout channel consists of a charge sensitive amplifier (CSA), an analog-to-digital converter (ADC), and a digital shaper. The gain of the CSA is 2 mV/fC, and the output signal of CSA will be digitized by the ADC with a 100 MHz sampling frequency. The digital shaper outputs a quasi-Gaussian signal with a shaping time of approximately 600 ns and a dead time of approximately 1.25 µs, indicating that each readout channel can tolerate a high counting rate of 800 kHz. If two or more signals generated in an electronic channel within 1.25 µs, only the first signal could be recorded. And finally, the valid data is eventually uploaded to the chassis controller via the PCIe bus.

3. Detector performance test and evaluation

    3.1 Two-dimensional imaging reconstruction algorithm

    The 2-D imaging reconstruction process has 4 steps. First, the data files from the 12 sub boards are read and resorted according to the timestamps of the X and Y channels, respectively, and then recorded as the complete data file. Then, the signals generated by different particles are separated according to the timestamp. The third step is reconstructing the hit point of each particle using the last detected signal in an event as the incident position. The last step is traversing the complete data file to make the 2-D imaging and calculating the interested distributions, such as the single channel signal amplitude, the event energy distribution, and the relationship between energy deposition and fired channel number in an event.

    3.2 Counting rate capability test with X-ray

    Many MPGD contains the resistive layer to suppress sparking, and it also causes a voltage drop

on the resistive layer [15]. Under high counting rate, the voltage drop would be tens of volt, leading to a significant voltage decrease in the electron multiplying region of MPGD and a much lower detector gain. With a decreased gain, the detector spectrum and stability would be various. To study the counting rate capability of the detector, a test with 8 mm-diameter X-ray is carried out. Figure 4 presents the simulated and experiment measured detector gain curves, where $G_0$ is the detector gain under low counting rate. The simulation method is based on MATLAB matrix calculation, considering both the flowing current effect and the accumulated charge effect caused by the resistive layer. The previous research describes the principle and demonstrates the effectiveness of the simulation method [15]. In this detector, the Ge-coated resistive layer has a thickness of 300 nm, corresponding to a surface resistance of approximately 100 MΩ/□. This result indicates that the measured detector gain-counting rate relationship is close to simulated curves with a surface resistance of 100 to 120 MΩ/□, showing a very good match. In this measurement, the 8.1 keV Cu X-ray can deposit all the energy in the 5 mm gas gap. In the real BNCT flux imaging, the heavy charged particle after n($^6$Li, $^3$H)$^4$He reaction, the recoil nucleus, and the MeV level gamma ray can penetrate the gas gap, and deposit more than 8.1 keV energy in the gas. Thus, the detector gain under low counting rate is around 3000 according to this measurement, can meet the demands of neutron and gamma measurement for the BNCT flux imaging.

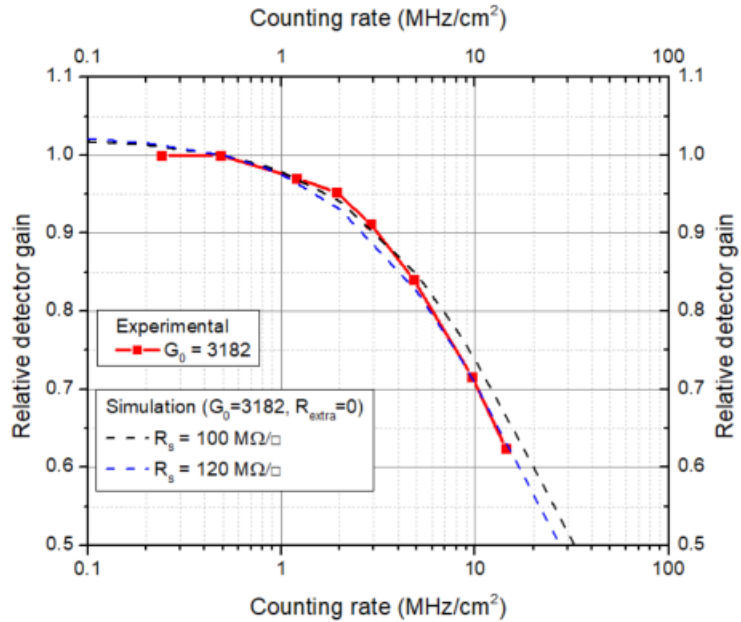

Fig. 4 Relationship between detector gain and counting rate with Φ8 mm X-ray irradiation area.

3.3 Spatial resolution evaluation

To measure the detector spatial resolution, a semicircle $^6$LiF film is attached to the drift cathode with an inclination angle of 45 degrees, and the detector is irradiated by an Am-Be source, as shown in Fig. 5 (a). Figure 5 (b) shows the reconstructed 2-D imaging of the semicircle neutron converter, and its edge indicates the detector spatial resolution. Rotating the imaging by 45 degrees and accumulating the hits along the Y axis, the edge could be fitted by a sigmoid function [17], as shown in Fig. 5 (c, d). Equation (2) shows the sigmoid function, where $A_{tot}$ is the height of the rising edge, the $X_{half}$ is the middle point of the edge, the s is the shape parameter, and the Constant is the longitudinal difference between the lower and the upper edge. The fitted shaping parameter of the Sigmoid function is 0.596 mm, the spatial interval between 10% and 90% of the total altitude is 4.4 mm, indicating that the detector spatial resolution is approximately 1.4 mm (~3σ). The measurement method influences the spatial resolution. In the measurement, the straight edge of semi-circle film is set as 45 degrees, and the accumulated counting in X' axis is proportional to the area covered by the film, as shown in Fig. 6 (a). It obvious that this measurement cannot generate a single step edge if we counting the hits in each grid point along the diagonal direction, and leading to a resolution of approximately 0.7 to 1 mm, as shown in Fig. 6 (b) and (c). The calculated spatial resolution of the measurement, 1.4 mm, is the co-action of the edge shape and detector spatial resolution. Using the function of $\sigma_{measured} = \sqrt{\sigma_{edge}^2 + \sigma_{detector}^2}$ , the detector spatial resolution could be calibrated as ~1 to 1.2 mm. Figure 10 shows a very clear dot array in the beamline range, which is caused by the spacers of the micromegas with a diameter of 1 mm, indicating a good detector spatial resolution. This result also demonstrates that the 1.4 mm is a very conservative estimation result.

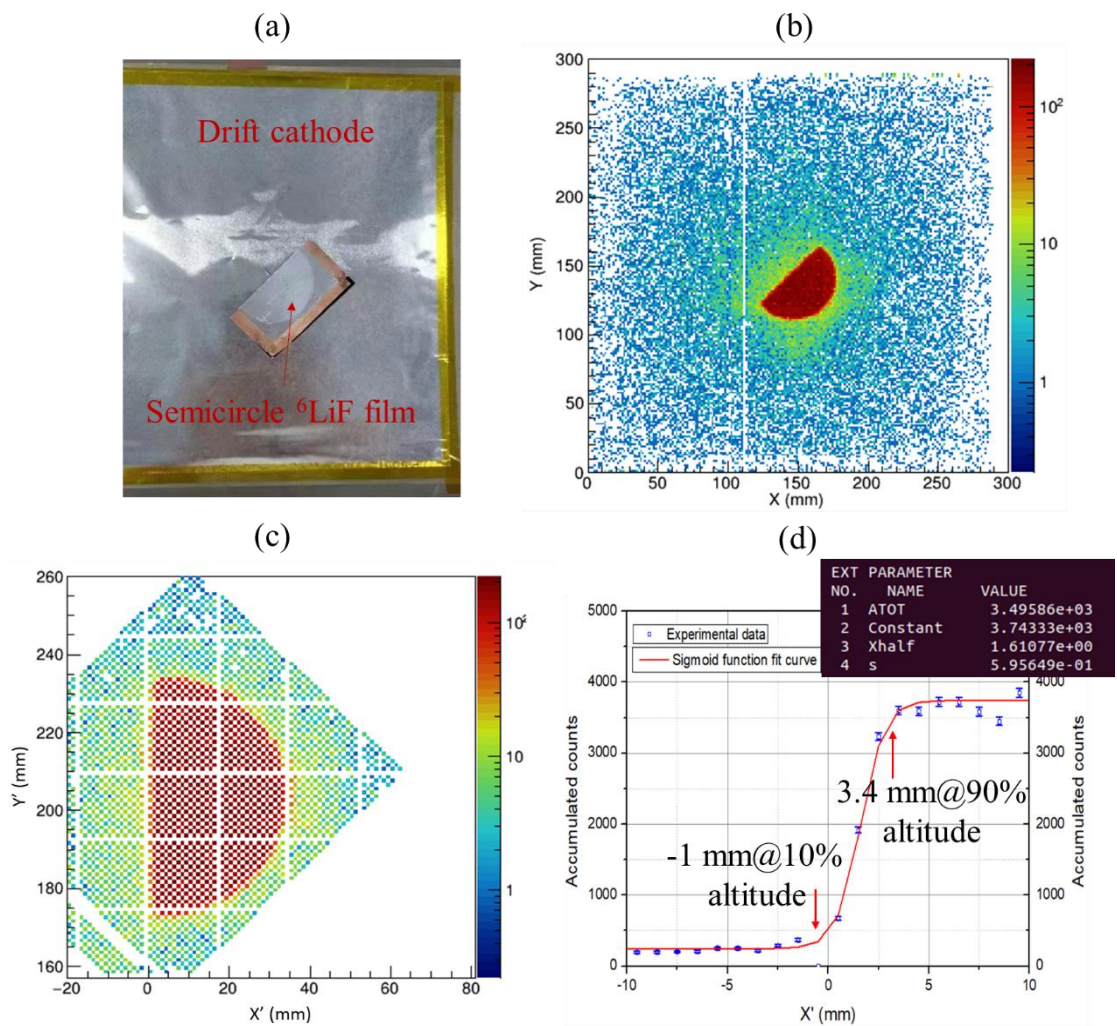

Fig. 5 (a) The semicircle $^6$LiF film attached to the drift cathode. (b) Reconstructed 2-D imaging of the semicircle film. (c) Rotated 2-D imaging. (d) The sigmoid function fitting of the left curve of the semicircle imaging.

$$A(x) = -\frac{A_{tot}}{1+\exp\left[\frac{(x-X_{half})}{s}\right]} + \text{Constant} \qquad \text{Equation (2)}$$

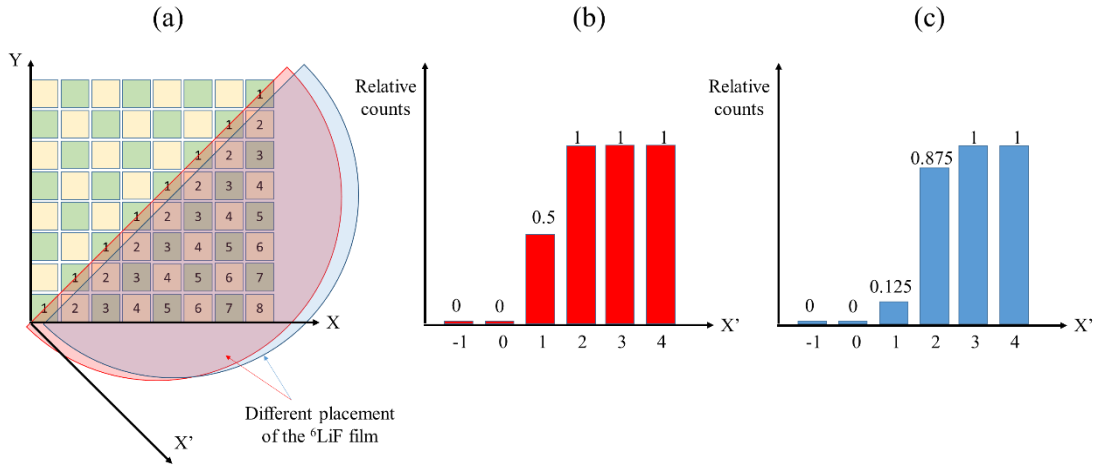

Fig. 6 (a) The schematic of semicircle $^6$LiF film placement. (b) The accumulated counts boundary formed by two step functions, corresponding to the red-colored $^6$LiF film placement in (a). (c) The accumulated counts boundary formed by three step functions, corresponding to the blue-colored $^6$LiF film placement in (a).

4. BNCT beam test and evaluation

    4.1 BNCT beam experiment layout

    Both the thermal and fast neutron detectors are tested in the in-hospital neutron irradiator BNCT beam at Beijing Capture Technology Corporation [13]. As shown in Fig. 7, the detector is positioned close to the beam outlet, and the center of the detector coincides with the beam center as much as possible. In the experiments, the thermal neutron detector is used to directly measure the thermal neutron component. Adding the 1 mm thick $^{nat}$Cd shield in front of the thermal neutron detector, the incident thermal neutron component could be suppressed by 99.9%, and the incident epithermal neutron component (0.5 to 10 eV) has an equivalent absorption probability of approximately 6%, according to the cross-section distribution of $^{113}$Cd [16]. Thus, it can be used to measure the epithermal neutron component. The fast neutron detector is used to measure the fast neutron component and gamma component with different working voltage settings.

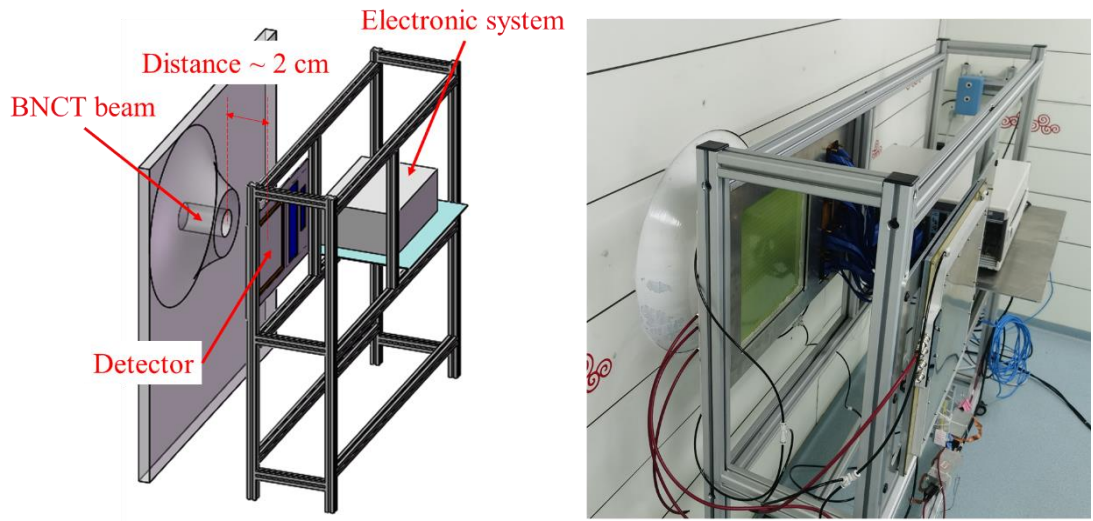

Fig. 7 (Left) Schematic of the BNCT experimental layout. (Right) Adjusting detector position in the measurement.

4.2 Two-dimensional imaging and neutron component evaluation

Figure 8 shows the relationship between energy deposition and fired channel number in an event (the E–N distribution), measured by the thermal neutron and fast neutron detectors, respectively. The fired channel number is the number of electronic channels that generate induced signal by an incident particle. In the thermal neutron detector, there are significant $^4$He and $^3$H particle distributions generated by the $^6Li(n,\alpha)^3H$ reaction. Between the two heavy charged particles, there is a fast neutron recoil ion distribution. The gamma signal is close to the left-down corner. In the fast neutron detector, the gamma ray can generate centrally distributed signals with low energy deposition and a large fired channel number, and the rest is the fast neutron recoil ion signal. Thus, according to the E–N distribution, signals generated by various components can be separated.

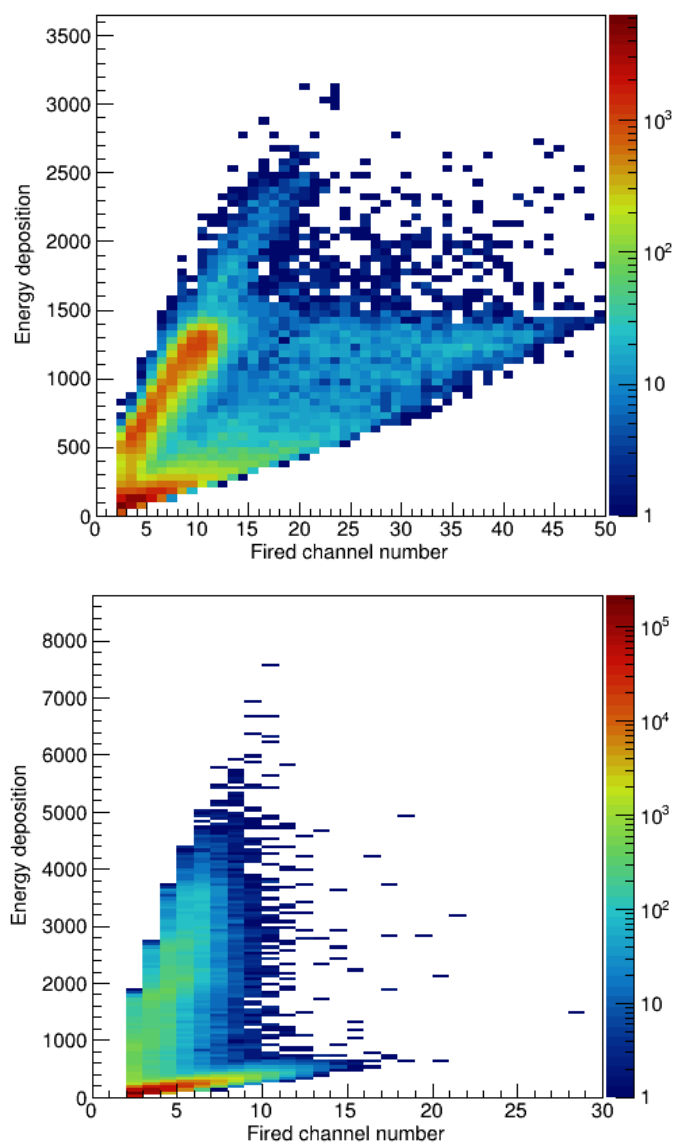

Fig. 8 The relation between energy deposition and fired channel number in an event in a thermal neutron beam. (Upper) The thermal neutron detector. (Lower) the fast neutron detector.

Figure 9 shows the reconstructed 2-D imaging with thermal neutron, epithermal neutron, fast neutron, and gamma ray components of the epithermal neutron beam. Figure. 10 shows the 2-D imaging of the thermal neutron beam. All the neutron components show a distinct circular pattern with a diameter of about 13 cm, slightly larger than the beam outlet diameter of 12 cm, probably due to the cone-shaped design of the neutron flux outlet and the non-collimated experimental measurement. Additionally, it should be noted that the signals generated by thermal neutrons and epithermal neutrons could not be separated because the generated heavy charged particles are the same. In the epithermal neutron beam measurement, the proportion of thermal neutron component is one order of magnitude lower than that of epithermal neutron component. However, the $^{nat}$LiF layer has significantly higher cross section with the thermal neutron, the detected thermal neutron hit number and epithermal neutron hit number are similar. In the thermal neutron beam, the thermal neutron component dominates and has larger cross section than the epithermal neutron component, resulting in a much higher counting rate and a better 2-D imaging performance. This also indicates the necessity of using $^{nat}$Cd shield in the epithermal neutron component measurements. The detector settings for this set of experiments are presented in Table. 1, indicating that the detector working voltage should be adjusted to meet the different requirements of various target particles, especially for the fast neutron and gamma components. Based on these data, the flux of different neutron components can be evaluated. With the help of the energy deposition and fired channel number relationship (Fig. 8), the different component flux could be calculated by selecting a specific portion in the distribution and then counting the reconstructed hits. As mentioned in section 3.3, the absorption probability of 1 mm $^{nat}$Cd to thermal neutrons and epithermal neutrons is 99.9% and 6% (assuming uniform distributed energy spectra), respectively. Also, it can be calculated that the equivalent detection efficiency between thermal neutrons and epithermal neutrons by the 15 nm $^{nat}$LiF film is 14.5:1. Thus, the fraction of the epithermal neutron component in the beam can be calculated, as shown in Table. 2. It can be seen that the calculated thermal neutron component is about an order of magnitude higher than both epithermal neutron and fast neutron components in the thermal neutron beam, and the calculated epithermal neutron shows a major component in the epithermal neutron beam, indicating a well moderated flux quality. In the measurement, the detector working voltage is adjusted to evaluate the threshold influence and the error of the measured flux. Increasing the working voltage by 20 V at 5% reactor power, the detected thermal and epithermal neutron rate is changing from 0.98 kHz to 1.06 kHz, with an increase of 8%. The measured fast neutron rate also increases from 6.5 kHz to 7.5 kHz when the working voltage increasing by 20 V at 10% reactor power. This result indicates that the thermal and epithermal neutron flux has a better certainty, for the most of the $^3$H and $^4$He deposit tens to 300 keV energy in the 5 mm gas gap. For the fast neutron, the recoil nucleus has a various energy from tens of eV to several MeV, resulting in a larger influence by the threshold. To make a conservative estimation, the thermal neutron detector has a measurement error of approximately 8%, and the fast neutron detector has a measurement error of approximately 15%. This set of experiments demonstrates that the micromegas-based neutron detector could realize large area 2-D imaging and flux evaluation for different components of the BNCT neutron beams.

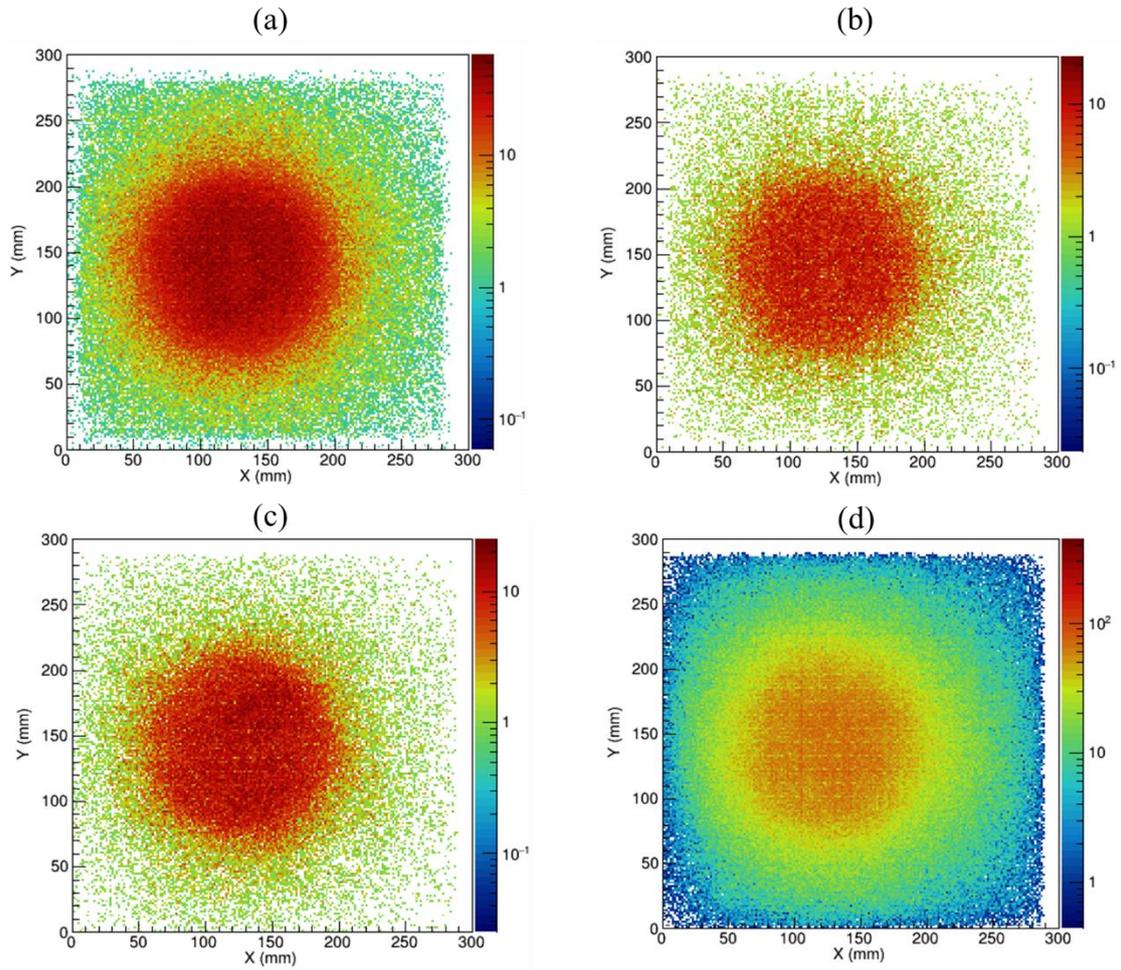

Fig. 9 Reconstructed 2-D imaging of the epithermal neutron beam. (a) Thermal and epithermal neutron components. (b) Epithermal neutron component. (c) Fast neutron component. (d) Gamma component.

Table. 1 Detector settings for various particle component measurements.

| Detector type | Target particle | $^{nat}$Cd shield | Micromegas working voltage (V) | Detector gain (approximately) |
|---|---|---|---|---|
| Thermal neutron detector | Thermal neutron | no | -390 | 100 |
| | Epithermal neutron | yes | -390 | 100 |
| Fast neutron detector | Fast neutron | no | -410 | 200 |
| | Gamma | no | -480 | 1300 |

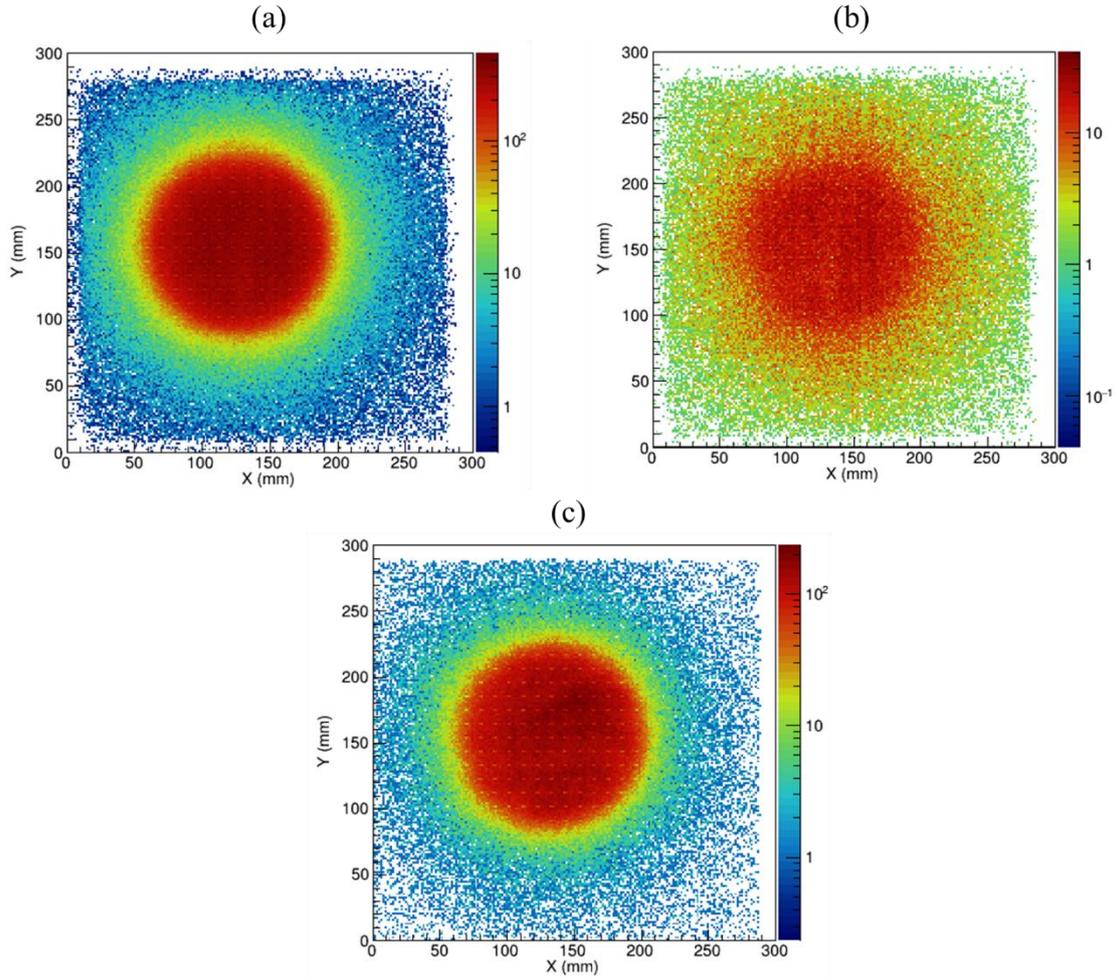

Fig. 10 Reconstructed 2-D imaging of the thermal neutron beam. (a) Thermal neutron component. (b) Epithermal neutron component. (c) Fast neutron component.

Table. 2 Neutron component flux calculation.

| Beam | Detected components | Total counting rate @10% power (Hz) | Equivalent detection efficiency | | | Calculated flux @10% power (Hz) |
| --- | --- | --- | --- | --- | --- | --- |
| | | | Thermal neutron | Epithermal neutron | Fast neutron | |
| Thermal neutron beam | Thermal neutron + epithermal neutron | 179 kHz | $6.6\times10^{-6}$ | $4.6\times10^{-7}$ | / | Thermal neutron: $2.7\times10^{10}$ |
| | Epithermal neutron | 0.58 kHz | 0 | $4.3\times10^{-7}$ | / | Epithermal neutron: $1.4\times10^{9}$ |
| | Fast neutron | 46 kHz | / | / | $1.0\times10^{-5}$ | Fast neutron: $4.5\times10^{9}$ |
| Epithermal neutron beam | Thermal neutron + epithermal neutron | 1.96 kHz | 6.6×10⁻⁶ | 4.6×10⁻⁷ | / | Thermal neutron: 1.0×10⁸ |
| | Epithermal neutron | 1.2 kHz | 0 | 4.3×10⁻⁷ | / | Epithermal neutron: 2.8×10⁹ |
| | Fast neutron | 9.57 kHz | / | / | 1.0×10⁻⁵ | Fast neutron: 9.4×10⁸ |

4.3 Counting rate linearity and highest counting rate evaluation

Figure 11 shows the detector single channel counting rates at various BNCT reactor powers, with and without the dead time calibration. The detector counting rate of the thermal beam is much higher (as seen in Table. 2), resulting in a more significant dead time calibration performance. It can be calculated that the linearity of the detector counting rates is below 1% when considering the 1.25 μs dead time of the single channel, indicating that both the thermal neutron and fast neutron detectors work well in the thermal or epithermal neutron beam.

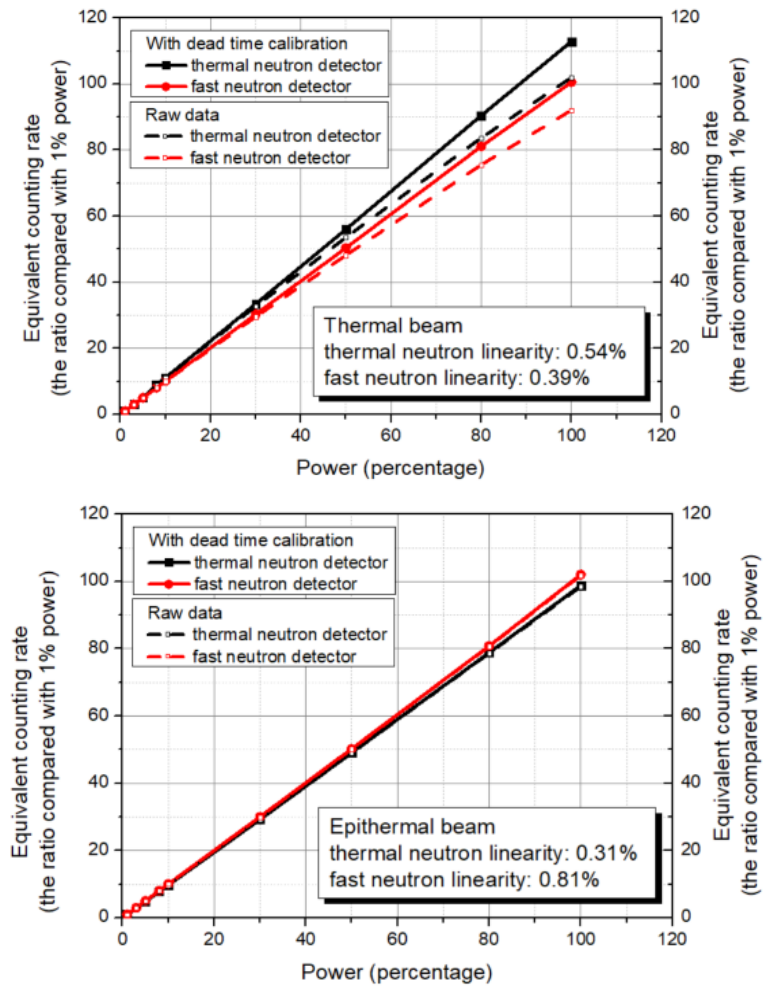

Fig. 11 (Upper) The counting rate linearity of the thermal neutron beam. (Lower) The counting rate linearity of the epithermal neutron beam.

When the thermal neutron detector measures the thermal neutron beam with 100% power, the detector has the highest counting rate. The measurement data indicate that the reconstructed particle counting rate is 8.79 kHz/cm$^2$, corresponding to the highest counting rate of 94 kHz/channel. The counting rate also varies with the different gas gap width. For example, the measured highest counting rate of the 3 mm gas gap thermal neutron detector is approximately 80 kHz/channel. Figure 12 presents the relationship between the detector gain and the counting rate of the two detectors in the thermal neutron beam. It is evident that the detector gain decrease can be suppressed below 10% at 100% power, indicating an acceptable stability of detector performance. In both X-ray test and BNCT measurement, the detected signal has an amplitude of hundreds of mV, indicating that the product of particle energy deposition and detector gain under low counting rate is close. Thus, the measured detector gain curve shows a similar gain performance in the counting rate range of [1, 10] kHz/cm$^2$ with the X-ray test result, as well as the simulation data.

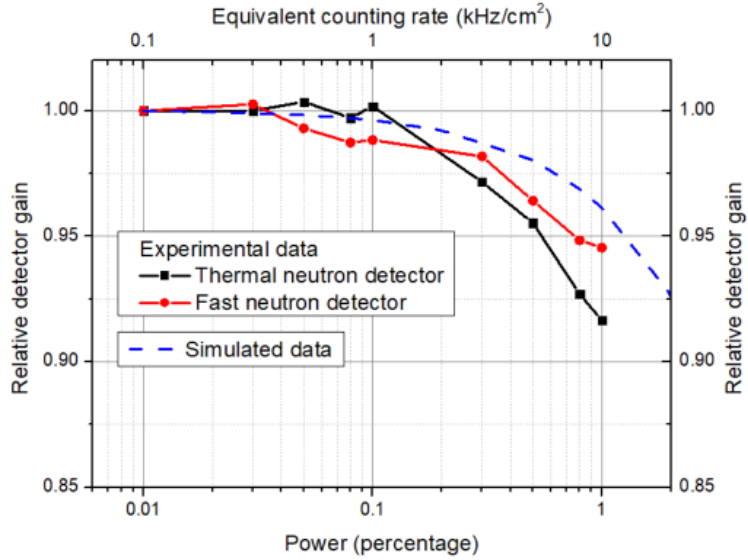

Fig. 12 Relationship between detector gain and counting rate with large irradiation area.

4. Discussion

In this detector design, the X and Y strips have a pitch of 1.5 mm, and the time window for X and Y signal matching is set as 100 ns. In the 100% power thermal neutron beam with a diameter of approximately 13 cm, the highest hit rate is 8.79 kHz/cm$^2$, indicating a hit rate of 1167 kHz in total. Thus, ghost hits may be a problem. With a 100 ns time window, the probability that one time window contains more than one particle track can be calculated as 11%, as shown in Equation (3). This may cause an incorrect reconstruction of the particle incident position, affecting the quality of 2D imaging. To address the problem, it is an effective way to make full use of the time information of the signal to reconstruct the most likely track, or optimize the time resolution of the system.

$$Total\ hit\ rate = 1167\ kHz$$
$$Time\ window = 100\ ns$$
$$\therefore t_0 = \frac{1}{1167k} = 857 ns$$

$$\therefore P = 1 - e^{-T_0/t_0} = 1 - e^{-100/857} = 11\% \qquad \text{Equation (3)}$$

5. Conclusion

Based on micromegas technology, we implement thermal neutron and fast neutron detector designs with a large area and high counting rate capability for beam monitoring and evaluation in BNCT. This detector has a compact structure design, with a detection area of 288 mm × 288 mm and 1.5 mm pitch 2-D readout strips. The experiments in IHNI-I BNCT thermal and epithermal neutron beams demonstrate that, with 15 nm thick $^{nat}$LiF neutron converter film and 1 mm thick $^{nat}$Cd shield, the detector can be used to measure the thermal and epithermal neutron components. Using Ar:$CO_2$=93:7 as the working gas, the detector can measure the fast neutron and gamma ray components. The experiment also proves that this detector can realize good 2-D imaging for the above particle components and achieve the highest counting rate of 94 kHz/channel. The Am-Be neutron source experiment also indicates that this detector has a spatial resolution of approximately 1.4 mm. These experimental results demonstrate that this micromegas-based neutron detector can meet the requirements of BNCT beam monitoring and evaluation and has great development prospects.

Acknowledgements
We thank the Hefei Comprehensive National Science Center for its strong support. This work was partially performed at the University of Science and Technology of China (USTC) Center for Micro and Nanoscale Research and Fabrication, and we thank Yu Wei for his help in the nanofabrication steps of the Ge coating for the resistive anode.

Funding
This work was supported by the Science and Technology on Metrology and Calibration Laboratory (Grant No. JLJK2021001C003), Fundamental Research Funds for the Central Universities, China (Grant No. WK2360000011 and WK2360000010), Guangdong Basic and Applied Basic Research Foundation (Grant No. 2022B1515120032).